# QUASI-ORTHOGONAL SPACE-TIME-FREQUENCY TRELLIS CODES FOR MIMO-OFDM SYSTEMS

Robinson Ebi Elias. J and Rajesh. R

School Of Electronics Engineering, VIT University

Vellore-632014, Tamilnadu, India

vitrobinson@gmail.com;rajesh@vit.ac.in

## ABSTRACT

*The main objective of this project is to design the full-rate Space-Time-Frequency Trellis code (STFTC), which is based on Quasi-Orthogonal designs for Multiple-Input Multiple-Output (MIMO) Orthogonal Frequency Division Multiplexing (OFDM) systems. The proposed Quasi-Orthogonal Space-Time-Frequency Trellis code combines set partitioning and the structure of quasi-orthogonal space-frequency designs in a systematic way. In addition to multipath diversity and transmit diversity, the proposed code provides receive diversity, array gain, and achieve high-coding gain over a frequency selective fading channel. As simulation results demonstrate, the code outperforms the existing Quasi-Orthogonal Space-Time-Frequency Trellis codes in terms of frame error rate performance.*

## KEYWORDS

*MIMO-OFDM, receive diversity, quasi-orthogonal designs, trellis codes, space-time-frequency codes.*

## 1. INTRODUCTION

Spatial diversity is a popular diversity method for combating the effects of fading without the need to increase the bandwidth for the case of narrowband wireless communication, where the fading channel is frequency non-selective. Space diversity can be implemented in the form of transmit and/or receive diversity creating Multiple-Input Multiple-Output (MIMO) channels. In case of broadband wireless communications, where the fading channel is frequency selective, Orthogonal Frequency Division Multiplexing (OFDM) modulation can be used to transform the frequency-selective channel into set of parallel frequency non-selective channels. To take advantage of both MIMO systems and OFDM modulation, MIMO-OFDM systems have been proposed [1], which guarantees reliability and high spectral efficiency for next-generation wireless communication systems. There have been a lot of efforts in designing codes to obtain a high diversity for OFDM systems [2]-[6]. This is achieved by designing space-time codes [2], [4], space-frequency codes [3], and space-time-frequency codes [5], [6]. Space-Time-Frequency (STF) coding schemes can achieve a maximum diversity gain equal to the product of the number of transmit antennas $M_t$, receive antennas $M_r$, the number of propagation paths $L$, and the rank of the channel temporal correlation matrix $R$ $(M_t M_r L R)$.Author in [7] introduced Orthogonal Space-Time Block Codes (OSTB) for two transmit antennas, It is noteworthy that in the case of more than two transmit antennas the OSTBC can provide a rate of at most $\frac{3}{4}$ and we are not able to have rate-one transmission. Quasi-Orthogonal Space-Time Block Codes (QOSTBC) structures for quasi-static channels were first introduced [8] and [9]. Original QOSTBC designs provide rate-one codes and pair-wise Maximum Likelihood (ML) decoding but fail to achieve full diversity. Later on, improved quasi-orthogonal codes were proposed through constellation rotation [10]-[13]. In [14] uses multiple antennas in receiver which provides array gain and receive diversity gain. In many of these schemes, good performance and the frequency and spatial diversities are guaranteed by concatenating a Trellis Coded Modulation (TCM) with a space-time-frequency code. However, these proposals have several drawbacks such as not being optimized for MIMO-OFDM systems, having a low coding rate, a large number of trellis states and a low coding gain among others.

Then, motivated by a need for trellis codes with good performance, low number of trellis states and low decoding complexity, we propose a scheme called Quasi-Orthogonal Space-Time-Frequency Trellis Codes (QOSTFTCs), where we combine Quasi-Orthogonal-Space-Time-Frequency Block Code (QOSTFBC) [15] for a Frequency Selective Channel (FSC) with four taps, with a trellis code in a systematic way. Note that for this case ($L$=4), the QOSTFBC is related to the QOSTBC with eight transmit antennas, with the difference that the QOSTFBC is implemented as a block diagonal quasi-orthogonal structure to take advantage of coding across the space, time, frequency dimensions. There has been no previous work on QOSTF trellis codes for four transmit antennas with parallel transitions in the trellis structure, such that both the multipath diversity and coding gain can be achieved despite parallel transitions. Furthermore, the QOSTF trellis schemes proposed in this letter are based on the design criteria derived in [18], [17] with appropriate arrangements. Although the code from [17] exploits the multipath diversity and transmit diversity gains available in the MIMO-OFDM channel, it does not provide additional coding gain, array gain and receive diversity as our proposed code do for four transmit and two receive antennas.

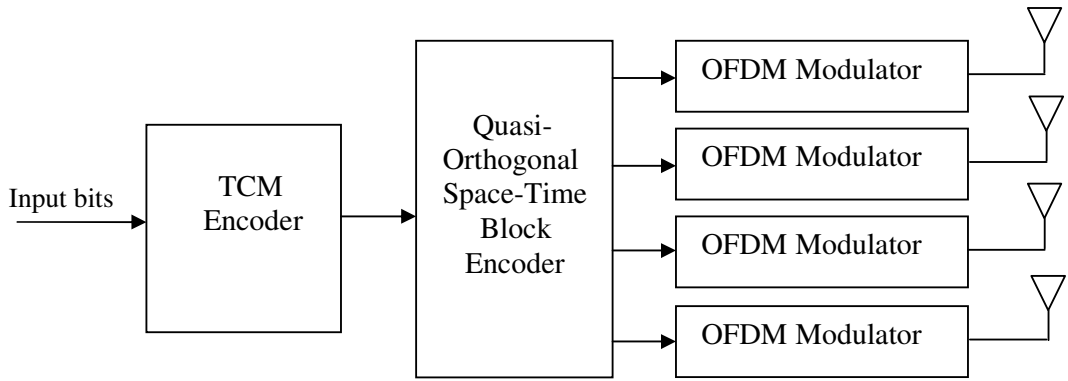

Figure 1. Transmitter block diagram for QOSTFTC

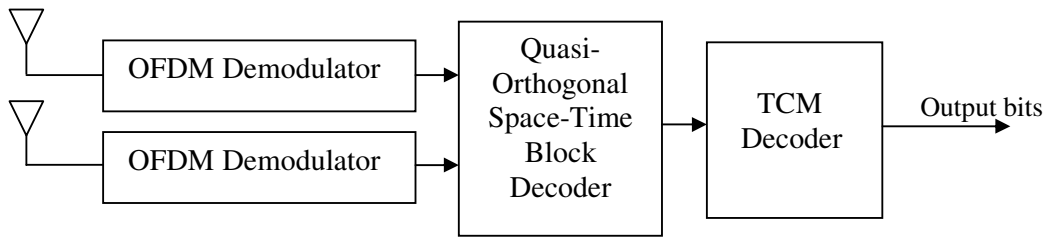

Figure 2. Receiver block diagram for QOSTFTC

## 2. SYSTEM MODEL

The transmitter and receiver block diagram of proposed QOSTFTC for MIMO-OFDM system with four transmit antennas ($M_t$=4) and two receive antenna ($Mr$=2) is given in Figures 1 and 2. Each transmit and receive antenna employs an OFDM modulator and demodulator with $N$ subcarriers. Assume that the receiver has perfect channel knowledge while the transmitter does not know the channel. There is no spatial fading correlation exists in between antennas throughout this work. And also assume that the channel impulse response (CIR) between the transmit antenna $p$ and receive antenna $q$ has $L$ independent delay paths on each OFDM symbol and an arbitrary power delay profile is given by

$$h_{p,q}(t) = \sum_{l=0}^{L=1} \alpha_{p,q}(l)\delta(t - \tau_l) \qquad (1)$$

Where $\tau_l$ represents the $l^{th}$ path delay and $\alpha_{p,q}(l)$ are the fading coefficients at delay $\tau_l$. It is assumed that all channels have the same power-delay profile. Note that each $\alpha_{p,q}(l)$ is a zero mean complex Gaussian random variable with a variance of $\frac{\sigma_l^2}{4}$ on each dimension. For normalization purposes, assume that $\sum_{l=0}^{L-1} \sigma_l^2 = 1$ in each transmit-receive link. It is necessary to append a cyclic prefix to each OFDM symbol to avoid Inter Symbol Interference (ISI) which is caused by the multipath delay of the channel, the channel frequency response (CFR), i.e. the fading coefficient for the $n$th subcarrier between transmit antenna $p$ and receive antenna $q$ is given by

$$H_{p,q}(n) = \sum_{l=0}^{L-1} \alpha_{p,q}(l)\, e^{-j2\pi n\Delta_f \tau_l} \qquad (2)$$

Where $\Delta_f$ is the inter subcarrier spacing, $\tau_l = l\, T_s$ is the $l^{th}$ path delay and $T_s = \frac{1}{N\Delta_f}$ is the sampling interval of the OFDM system. A space –frequency codeword for four transmit antennas transmitted at $l^{th}$ OFDM symbol period can be represented by,

$$\mathbf{C}_{SF}^t = [c_1^t(n)\; c_2^t(n)\; c_3^t(n)\; c_4^t(n)] \; \epsilon \; \mathbb{C}^{N \times M_t}$$

Where $c_p^t(n)$ is the complex data transmitted by the $p^{th}$ transmit antenna at the $n^{th}$ subcarrier, n=0,...,N-1. Moreover, $\mathbf{C}_{SF}^t$ satisfies the power constraint $E\|\mathbf{C}_{SF}^t\|_F^2 = N$. A space-time-frequency codeword has an additional dimension of time added to the above space-frequency codeword. In general we can express a STF codeword transmitted during $t^{th}$ OFDM symbol by

$$\mathbf{C}_{STF} = [\mathbf{C}_{SF}^t \; \mathbf{C}_{SF}^{t+1} \; \mathbf{C}_{SF}^{t+2} \; \mathbf{C}_{SF}^{t+3}] \; \epsilon \; \mathbb{C}^{N \times 4 \times 4}$$

At receiver, after matched filtering, removing the cyclic prefix and applying the fast Fourier transform (FFT) on frequency tones, the received signal at receive antenna $q$ at the $n^{th}$ subcarrier during the $t^{th}$ OFDM symbol duration is given by

$$Y_q^t(n) = \sum_{i=1}^{4} c_p^t H_{p,q}^t + N_q^t(n) \qquad (3)$$

Where $q=1,\ldots,M_r$, and $N_q^t(n)$ is a zero mean circularly symmetric Gaussian noise term, with zero-mean and variance $N_o$ at $t^{th}$ symbol period.

## 3. DESIGN CRITERIA

In this section, Discuss the design criteria for the proposed Quasi-Orthogonal Space-Time-Frequency Trellis codes (QOSTFTCs) according to the criteria derived in [18], [17] with appropriate adjustment. Let $Z$ be the frame length and $C^z$ be the branch output at the $z^{th}$ coding step of the trellis encoder. Here the fading is quasi-static over four OFDM symbols, i.e. $R=1$. Let us consider that $C^z$ is a QOSTB codeword in [11] for four complex symbols $c_1\ c_2\ c_3\ c_4$.

Also , let $\bar{\mathbf{C}} = \begin{bmatrix} \mathbf{c}_1 & \mathbf{c}_2 & \mathbf{c}_3 & \mathbf{c}_4 \\ -\mathbf{c}_2^* & \mathbf{c}_1^* & -\mathbf{c}_4^* & \mathbf{c}_3^* \\ -\mathbf{c}_3^* & -\mathbf{c}_4^* & \mathbf{c}_1^* & \mathbf{c}_2^* \\ \mathbf{c}_4 & -\mathbf{c}_3 & -\mathbf{c}_2 & \mathbf{c}_1 \end{bmatrix}$ be the transmitted coded sequence such that , at the first symbol period, the OFDM symbol $\mathbf{c}_1 = (c_1^1 c_1^2 c_1^3 c_1^4 \ldots c_1^Z)^T$ is transmitted from the first antenna, the symbol sent to the second antenna is $\mathbf{c}_2 = (c_2^1 c_2^2 c_2^3 c_2^4 \ldots c_2^Z)^T$ ,the symbol sent to third antenna is $\mathbf{c}_3 = (c_3^1 c_3^2 c_3^3 c_3^4 \ldots c_3^Z)^T$ ,and the symbol sent to the fourth antenna is $\mathbf{c}_4 = (c_4^1 c_4^2 c_4^3 c_4^4 \ldots c_4^Z)^T$ .In the second symbol period, the OFDM symbol $-\mathbf{c}_2^*$ is transmitted from the first antenna, the symbol transmitted from the second antenna is $\mathbf{c}_1^*$, the symbol transmitted from the third antenna is $-\mathbf{c}_4^*$, and the symbol transmitted from the fourth antenna is $\mathbf{c}_3^*$.In the third symbol period, the OFDM symbol $-\mathbf{c}_3^*$ is transmitted from the first antenna, the symbol transmitted from the second antenna is $-\mathbf{c}_4^*$, the symbol transmitted from the third antenna is $\mathbf{c}_1^*$,and the symbol transmitted from the fourth antenna is $\mathbf{c}_2^*$.In the fourth symbol period, the OFDM symbol $\mathbf{c}_4$ is transmitted from the first antenna, the symbol transmitted from the second antenna is $-\mathbf{c}_3$,the symbol transmitted from the third antenna is $-\mathbf{c}_2$,and the symbol transmitted from the fourth antenna is $\mathbf{c}_1$. At the receiver a maximum likelihood decoder might decide erroneously in favour of the coded sequence.

$$\bar{\mathbf{E}} = \begin{bmatrix} \mathbf{e}_1 & \mathbf{e}_2 & \mathbf{e}_3 & \mathbf{e}_4 \\ -\mathbf{e}_2^* & \mathbf{e}_1^* & -\mathbf{e}_4^* & \mathbf{e}_3^* \\ -\mathbf{e}_3^* & -\mathbf{e}_4^* & \mathbf{e}_1^* & \mathbf{e}_2^* \\ \mathbf{e}_4 & -\mathbf{e}_3 & -\mathbf{e}_2 & \mathbf{e}_1 \end{bmatrix}.$$

### 3.1. Diversity

The diversity order varies from $rM_r$ to $\delta_H M_r$, when the branch output is a symbol vector, as derived in [18], where $r$ and $\delta_H$ are the minimum rank and minimum symbol Hamming distance over all pairs of distinct coded sequence, respectively. Moreover, in order to achieve the maximum diversity $(M_t M_r L)$, it is necessary condition that $\delta_H \geq M_t L$. Let $\mathbf{D}^z = \mathbf{C}^z - \mathbf{E}^z$ be a branch difference matrix between $\mathbf{C}^z$ and $\mathbf{E}^z$, where $\mathbf{C}^z$ and $\mathbf{E}^z$ denote the $z^{th}$ codeword in coded sequence $\bar{\mathbf{C}}\ and\ \bar{\mathbf{E}}$, respectively. A codeword distance matrix is defined as $\mathbf{A}^z = (\mathbf{D}^z)^H \mathbf{D}^z$.Next we define $\rho(\bar{\mathbf{C}}, \bar{\mathbf{E}})$ as the set of instances $1 \leq z \leq Z$ at which $\mathbf{C}^z = \mathbf{E}^z$ and $\delta_H$ as the number of elements in $\rho(\bar{\mathbf{C}}, \bar{\mathbf{E}})$. If $\mathbf{A}^z$ is a rank-four matrix for all $z \ \epsilon \ \rho(\bar{\mathbf{C}}, \bar{\mathbf{E}})$, it can be shown that the diversity ranges from $4M_r$ to $4M_r\delta_H$, and the maximum achievable diversity order of the proposed QOSTFTCs is $4M_r \cdot min\ (\delta_H, L)$ over any Frequency Selective Channel (FSC) with four transmit antennas and $L$ independent taps.

## 3.2. Coding Gain distance

To maximize the coding gain the distance criterion derived in [18] can be rewritten as the maximization of the minimum product of the coding gain distance $CGD = \det(\sum_z \mathbf{A}^z)$ and the modified product distance $MPD = \prod_z (1 + \|\mathbf{D}^z\|_F^2)$, where $z \in \rho(\overline{\mathbf{C}}, \overline{\mathbf{E}})$ as given in [17].

## 3.3. Optimal Rotation Angle

The Optimum rotation angle, $\phi$ are determined such that the coding gain is maximized and achieves full diversity and has pairwise maximum likelihood decoding. Based on [15] the optimum rotation angle for this code, for MPSK constellation is $\pi/M$ (for M even) and $\pi/2$ (for M odd).

Consequently, the following design steps are proposed:

a) Perform set partitioning for the available codewords. The set partitioning metric is the product *CGD. MPD* over all possible pairs of distinct codewords.
b) Expand the available codewords constellation as necessary to design full-rate QOSTFTCs.
c) Codewords that do not belong to the same codewords constellation are assigned to different states. Assign codewords diverging (or merging to) into a state such that $\mathbf{A}^z$ must have full rank, and all pairs of codewords diverging from or merging to a state must be separated by the largest product *CGD.MPD*.
d) In order to achieve the multipath diversity provided by the channel $\delta_H \geq L$ must be satisfied. It can be shown that the coding gain will increase when $\delta_H$ is increased.

Note that the design criteria of proposed code do not need any knowledge of the channel delay profiles. In order to eliminate the dependence on the channel delay profiles, it is common to use an interleaver between a trellis encoder and an OFDM modulator to achieve reasonable robust code performance [4].

# 4. QOSTFT CODES: STRUCTURE AND DESIGN

In the section, a code has been proposed using the above design criteria, to achieve rate-one, high coding gain, multipath diversity and receive diversity.

## 4.1. Quasi-Orthogonal STF Block codes

Consider a multipath channel described in system model where $M_t = 4$ transmit antennas. Assume a temporal diversity of $R$ is desired; therefore we spread our codeword across $R$ OFDM symbol durations. We choose a generalized QOSTBC code given by [19], corresponding to $4LR$ transmit antennas to build our QOSTF code. The codeword transmitted during the $t^{\text{th}}$ OFDM symbol duration is given by

$$\mathbf{C}_{STF}^t = \left[ \mathbf{G}_t^{1^T} \ \mathbf{G}_t^{2^T} \ \mathbf{G}_t^{3^T} \ \mathbf{G}_t^{4^T} \ \dots \ \mathbf{G}_t^{m^T} \ \dots \right]^T \in \mathbb{C}^{N \, X \, 4} \qquad (4)$$

Where $t \in \{1, \dots, R\}$ and for a block index of $m \in \left\{1, \dots, \left\lfloor \frac{N}{4L} \right\rfloor\right\}$, and we rearrange $\mathbf{G}_t^m$ in [15] with appropriate adjustments. In general, for larger temporal diversity advantage $R$, one can spread the codewords across an arbitrary number of OFDM blocks but there is a delay of $R$ OFDM symbols associated with the decoding process.

## 4.2. Quasi-Orthogonal STF Trellis codes

We propose a high-coding gain QOSTFTCs for four transmit antennas. For the system with a large number of transmit antennas and/or high order constellation modulation, it is very difficult to prevent parallel trellis transitions from happening.

### 4.2.1. Codeword structure

Let us assume a 4-ray channel model, Then, we rearrange the general class of QOSTFTCs given by [17] in the codeword matrix is

$$\mathbf{G}^z = \frac{1}{\sqrt{4}} \begin{pmatrix} x_1 + \tilde{x}_5 & x_2 + \tilde{x}_6 & x_3 + \tilde{x}_7 & x_4 + \tilde{x}_8 \\ -(x_2 + \tilde{x}_6)^* & (x_1 + \tilde{x}_5)^* & -(x_4 + \tilde{x}_8)^* & (x_3 + \tilde{x}_7)^* \\ -(x_3 + \tilde{x}_7)^* & -(x_4 + \tilde{x}_8)^* & (x_1 + \tilde{x}_5)^* & (x_2 + \tilde{x}_6)^* \\ x_4 + \tilde{x}_8 & -(x_3 + \tilde{x}_7) & -(x_2 + \tilde{x}_6) & (x_1 + \tilde{x}_5) \\ x_1 - \tilde{x}_5 & x_2 - \tilde{x}_6 & x_3 - \tilde{x}_7 & x_4 - \tilde{x}_8 \\ -(x_2 - \tilde{x}_6)^* & (x_1 - \tilde{x}_5)^* & -(x_4 - \tilde{x}_8)^* & (x_3 - \tilde{x}_7)^* \\ -(x_3 - \tilde{x}_7)^* & -(x_4 - \tilde{x}_8)^* & (x_1 - \tilde{x}_5)^* & (x_2 - \tilde{x}_6)^* \\ x_4 - \tilde{x}_8 & -(x_3 - \tilde{x}_7) & -(x_2 - \tilde{x}_6) & (x_1 - \tilde{x}_5) \end{pmatrix} \epsilon \, \mathbb{C}^{8 \times 4}$$

Where space goes horizontally, $\{x_1, x_2, x_3, x_4\}$ belong to a M-PSK constellation $A$ and $\{\tilde{x}_5, \tilde{x}_6, \tilde{x}_7, \tilde{x}_8\}$ belong to the rotated constellation $Ae^{j\phi}$. The optimum rotation is $\phi = \pi / M$ since it provides the maximum coding gain for the code in [17].

### 4.2.2. Set partitioning

Let $\mathbf{G}_1^z(x_1, x_2, x_3, x_4, \tilde{x}_5, \tilde{x}_6, \tilde{x}_7, \tilde{x}_8)$ and $\mathbf{G}_2^z(y_1, y_2, y_3, y_4, \tilde{y}_5, y_6, \tilde{y}_7, \tilde{y}_8)$ be the two codewords as defined in [16] and [17], full diversity is achieved if the CGD given as $\mathrm{CGD}(\mathbf{G}_1^z, \mathbf{G}_2^z) = \frac{1}{8}\left\{ \sum_{p=1}^{4} |(x_p - y_p) + (\tilde{x}_{p+4} - \tilde{y}_{p+4})|^2 + |(x_p - y_p) - (\tilde{x}_{p+4} - \tilde{y}_{p+4})|^2 \right\}$ is not zero. Furthermore, the minimal product *CGD. MPD* between codewords at each level of an optimal set partitioning must be maximum. We use the set partitioning given in [17] [16] with proper modification.

### 4.2.3. QOSTF Trellis code design

Due to symmetry, if $(x_1, x_2, x_3, x_4) \, \epsilon \, Ae^{j\phi}$ and $(\tilde{x}_5, \tilde{x}_6, \tilde{x}_7, \tilde{x}_8) \, \epsilon \, A$ in the codeword structure results in a code with similar properties and the full diversity is still achieved. This will give us a new degree of freedom and additional constellation matrices to pick from. In order to expand the constellation of matrices, let $\phi_1$, $\phi_2$, $\phi_3$ and $\phi_4$ be the rotation angles for the symbols $\{x_1, x_2, x_3, x_4\}$ and $\{\tilde{x}_5, \tilde{x}_6, \tilde{x}_7, \tilde{x}_8\}$, respectively. Then we set $\left(\phi_1 = 0, \phi_2 = 0, \phi_3 = \frac{\pi}{4}, \phi_4 = \frac{\pi}{4}\right)$ or $\left(\phi_1 = \frac{\pi}{4}, \phi_2 = \frac{\pi}{4}, \phi_3 = 0, \phi_4 = 0\right)$ with QPSK. We use a similar systematic design method given in [17] with proper modifications to assign the subsets in the proposed 4-state QOSTFTC.

### 4.3. QOSTFT code design for receive diversity

The Orthogonality of the subspaces of the generator matrix results in the possibility of decoding pairs of symbols independently. To simplify the complexity of decoding process by combining the set partitioning and separate decoding of the inner QOSTFTCs; furthermore it allows pairwise Maximum Likelihood (ML) decoding using the Viterbi algorithm. The receiver block diagram of proposed QOSTFTCs is shown in the Figure 2. We have $M_r = 2$ receive antennas, so we can use Maximum Ratio Combining (MRC) for the ML decoding with more than one receive antenna. Therefore we can write the cost function for only one receive antenna and add the correct summation in front of it to achieve the ML decoding for the general case of $M_r$ receive antennas as given in [20], we call this maximum ratio combining. Because of using two receive antennas, it can be shown that the diversity ranges from $8$ to $8$ $\delta_H$ and the maximum achievable diversity order of the proposed QOSTFTCs is $8 \times \min(\delta_H, L)$ over any FSC with four transmit; two receive antennas and $L$ independent paths. Also provides array gain where the receiver has perfect channel knowledge.

## 5. SIMULATION RESULTS

In addition to theoretical analysis, we present simulation to investigate the performance of our designs in a MIMO-OFDM system equipped with ($M_t$=4) four transmit antennas and ($M_r$=1,2) one and two receive antennas; each OFDM modulator utilizes 64 subcarriers with a total bandwidth of 1MHz,and the cyclic prefix length is long enough to combat ISI. We assume that the average symbol power per transmit antenna is $E_s = \frac{1}{M_T}$ and the noise variance is $\frac{1}{SNR}$. Assume that the channel is quasi-static over four symbol periods (a frame) and changes independently for each frame. The performance curves are described by means of frame error rate (FER) versus the receive SNR with a QPSK constellation. The proposed schemes are compared with the system using quasi-orthogonal space-time-frequency block (QOSTFBC) code) with two transmit antennas ($M_t$=2) presented in [17]. In order to observe the robustness of the proposed QOSTFTCs, a random interleaver is not applied. It can be seen from the slopes of the performance curves in Figure 3, that the proposed QOSTFBC for ($M_t$=4, $M_r$=1) antennas achieves full diversity order of 4 and outperforms the QOSTFBC in [17] by almost 4 dB. We can see from the FER curves in Figure 4, that because of the trellis encoding, the 4-state QOSTFTC achieves an additional coding gain of 3.6 dB despite of parallel transitions in the trellis structure for MIMO-OFDM system using ($M_t$=4, $M_r$=1) antennas. In Figure 5, the proposed 4-state QOSTFTC for ($M_t$=4,$M_r$=1) antennas outperforms the 16-state QOSTFTC for ($M_t$=2,$M_r$=1) antennas in [17] by 3.6 dB, and acheives high coding gain with reduced number of trellis states. In Figure 6, the performance of proposed QOSTFB code and QOSTFT code using ($M_t$=4, $M_r$=1) antennas is compared with proposed QOSTFB code and QOSTFT code using ($M_t$=4, $M_r$=2) antennas. It can be seen from the slopes of the performance curves in Figure 6, at a FER of $10^{-2}$ the QOSTFB code for two receive antenna ($M_r$=2) outperforms the QOSTFB code for one receive antenna ($M_r$=1) by 4.6 dB, and QOSTFT code for two receive antennas ($M_r$=2) ouperforms the QOSTFT code for one receive antenna ($M_r$=1) by 4 dB at a FER of $10^{-3}$, so the codes for two receive antennas acheives array gain and receive diversity in addition to transmit diversity, and the diversity order of 8. All of these observations are consistent with the properties of our proposed codes discussed in Section 4.

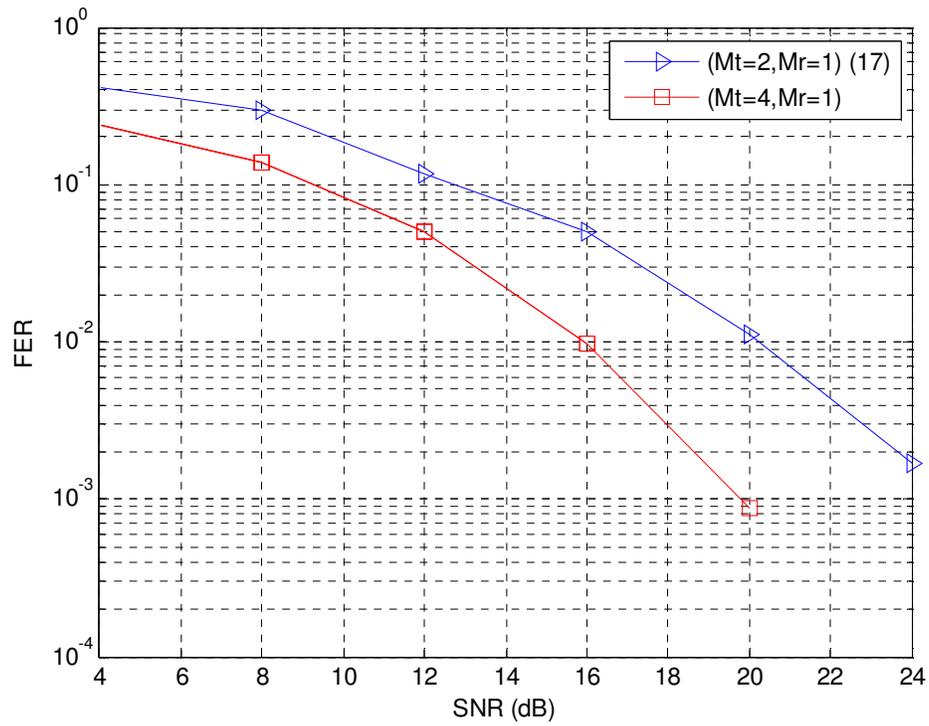

Figure 3. Performance of rate-one QOSTFB codes

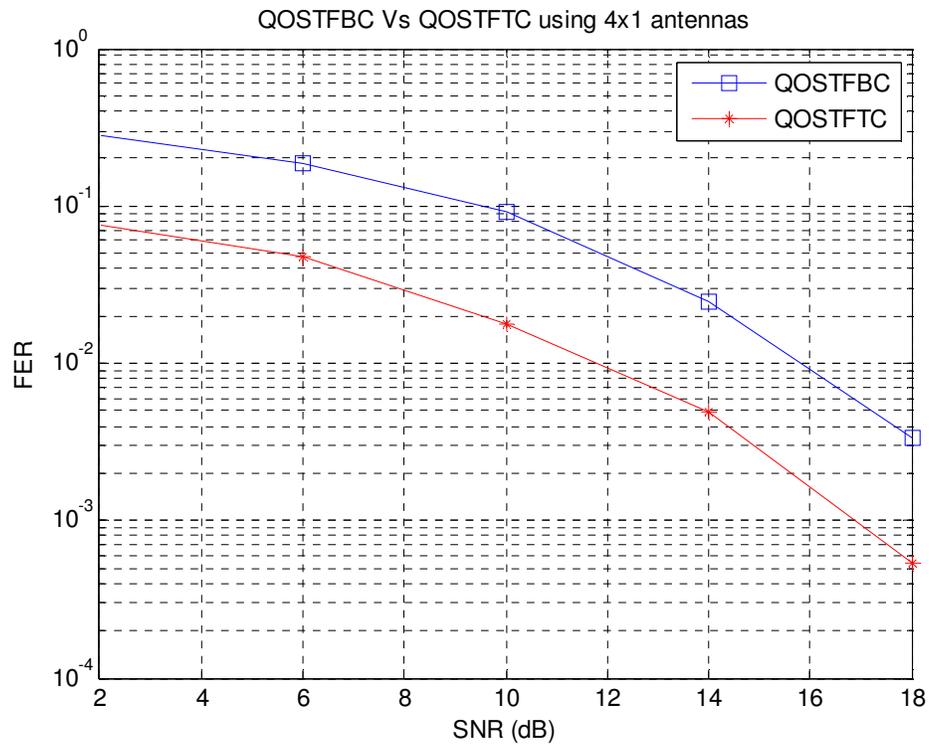

Figure 4. Performance of QOSTFTC with QOSTFBC

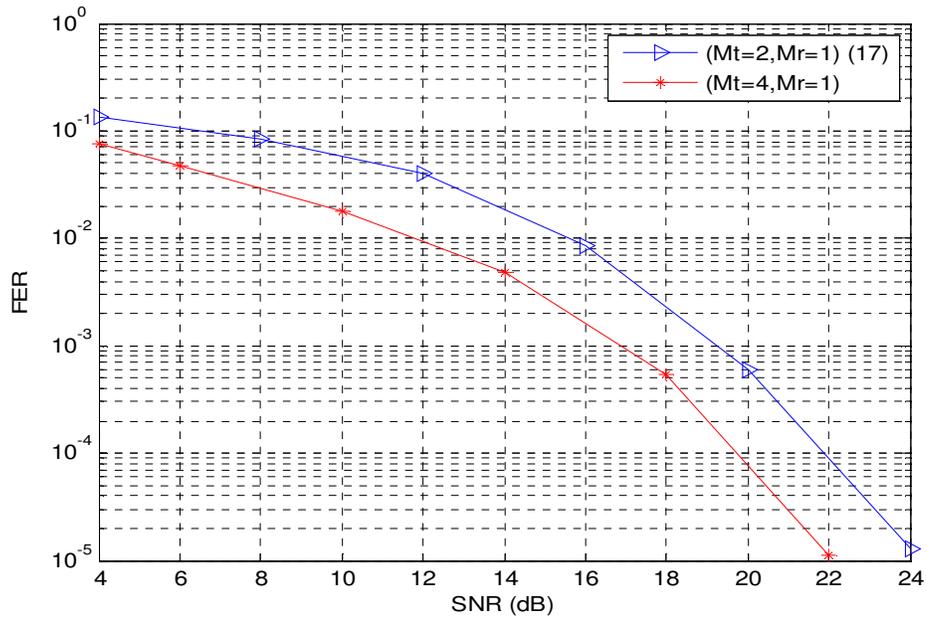

Figure 5.Performance of 4-stateQOSTFTC with 16-state QOSTFTC

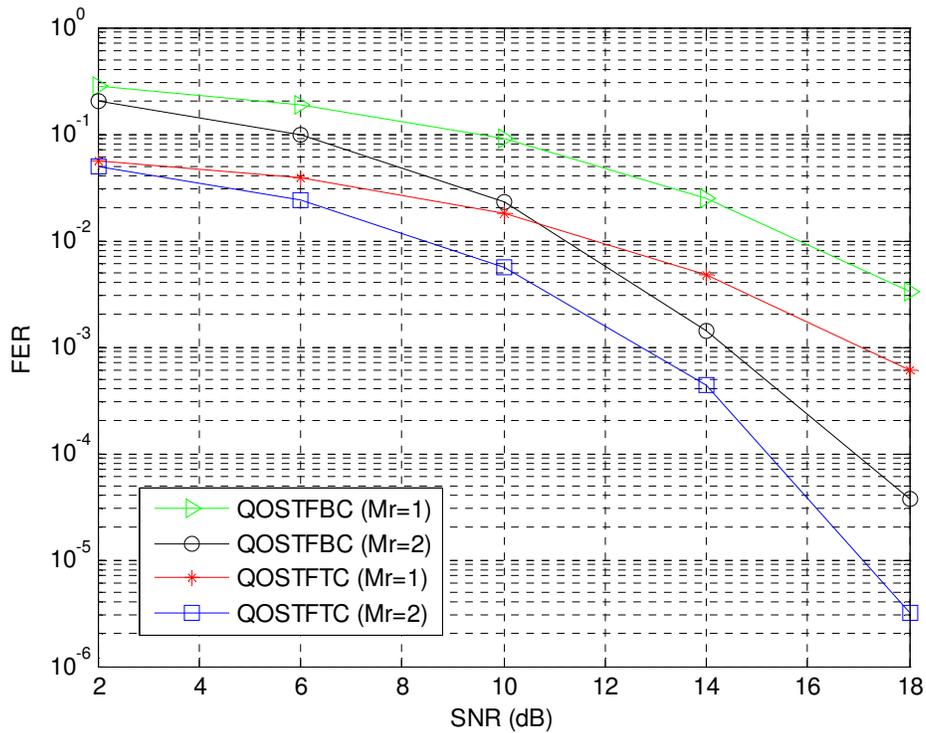

Figure 6. Performance of QOSTFBC and QOSTFTC using receive diversity

## 6. CONCLUSIONS

In this paper, we have carried out a design of rate-one, high-coding gain QOSTFTCs for MIMO-OFDM systems for four transmit and two receive antennas under a frequency selective fading channel. The proposed QOSTFTCs provides array gain, receive diversity in addition to multipath diversity and transmit diversity. Furthermore, as simulation suggest, without applying a channel interleaving strategy, our proposed QOSTFTCs outperforms the best available space-time-frequency trellis code in the literature. If the channel is quasi-static over four adjacent OFDM symbols, i.e. the channel stays constant for four adjacent OFDM symbols, there are no temporal diversity gains offered by the channel, and there is a delay of four OFDM symbols associated with the decoding procedure in all cases. However, the decoding complexity of the proposed QOSTFTCs is reduced.In order to achieve full diversity order offered by the channel, QOSTFTCs with larger $\delta_H$ are required. Note that we consider $L$=8 for the proposed designs,but it is straightforward to design similar codes for more than eight taps.